\documentclass[12pt]{iopart}

\usepackage{iopams}  
\usepackage{graphicx}
\usepackage{xcolor}
\begin{document}

\title{Novel neutron decay mode inside neutron stars}

\author{Wasif Husain$^{1}$ \& Anthony W. Thomas$^1$}

\address{$^1$ARC Centre of Excellence for Dark Matter Particle Physics and CSSM, Department of
Physics, University of Adelaide, SA 5005, Australia}

\ead{wasif.husain@adelaide.edu.au}
\vspace{10pt}
\begin{abstract}
We explore the suggestion that the neutron lifetime puzzle might be resolved by neutrons decaying into dark matter through the process, n $\rightarrow\chi\chi\chi$, with $\chi$ having a mass one third of that of the neutron. In particular, we examine the consequences of such a decay mode for the properties of neutron stars. Unlike an earlier suggested decay mode, in order to satisfy the constraints on neutron star mass and tidal deformability, there is no need for a strong repulsive force between the dark matter particles. This study suggests the possibility of having hot dark matter at the core of the neutron star and presents a possible mechanism of dark matter cooling, and examines the possible signal of neutrons decaying in this way inside the neutron star right after its birth.
\end{abstract}
\section{Introduction}
The observation of two different values for the neutron lifetime has been a mystery for some years. Neutrons in a beam show a longer lifetime than those trapped in a bottle ~\cite{RevModPhys.83.1173}. The lifetime of neutrons in a beam ~\cite{Otono:2016fsv,PhysRevLett.111.222501,Byrne_1996,PhysRevC.71.055502} is measured by counting the total number of protons produced, while the lifetime of neutrons trapped in a bottle is measured by counting the number of neutrons remaining after a known period of 
time~\cite{SEREBROV200572,PICHLMAIER2010221,PhysRevC.85.065503,ARZUMANOV201579}. The lifetime of neutrons in a beam appears to be approximately 8 seconds or 4$\sigma$ longer than that of neutrons in a bottle. A recent precise measurement of the lifetime of neutrons trapped in a bottle by the UCN$\tau$ Collaboration ~\cite{UCNt:2021pcg}, found the lifetime of the  neutron was 877.75 $\pm$ 0.28 (stat.) + 0.22 - 0.16 (syst.) seconds, which is similar to other recent measurements~\cite{scienceaan8895,PhysRevC.97.055503}.

A few years ago, Fornal and Grinstein proposed that the explanation for this discrepancy might be that neutrons have a small branching ratio for decay through an unknown dark channel (dark matter). This naturally explains why the lifetime of neutrons in a beam, with the decay process observed, appears longer than that for  neutrons in a bottle experiment, where the decay mode is 
unseen~\cite{PhysRevLett.120.191801,doi:10.1142/S0217732320300190,Rajendran_2021}. This new
decay channel 
\begin{equation}
    n\rightarrow  \chi + \phi,
    \label{eq:decay}
\end{equation}
must have a branching ratio around 1\%.
$\phi$ is an extremely light boson that goes undetected and $\chi$ is a dark matter fermion with baryon number one and mass in the narrow range 937.9 MeV $<$ $m_\chi$ $<$ 938.7 MeV \cite{PhysRevLett.120.191801,Serebrov:2018mva}. 

References ~\cite{2018_tanga,Agostini_2015} ruled out the possibility of $\phi$ being a photon. It was very quickly realized that this decay mode could be tested against the known properties of neutron 
stars~\cite{Motta:2018bil,Motta:2018rxp,PhysRevLett.121.061801,2018_sa}. Indeed, if the dark baryon is non-interacting the maximum neutron star mass is reduced to 0.7 M$_\odot$, which is unacceptable. Thus, for this explanation to survive the dark matter particles must be very strongly self-repulsive. 

Recently, an interesting alternative was proposed by Strumia~\cite{Strumia_2022}, in which it was suggested that the dark decay mode might involve neutrons decaying into three identical dark matter particles, each having baryon number 1/3. It was claimed that this model does not require dark matter to be self-repulsive to survive the neutron star maximum mass constraint. Here, we carry out a detailed examination of the consequences for neutron star properties of the model suggested in Ref.~\cite{Strumia_2022}.

The layout of this article is as follows. Section two explains the model used to describe the nuclear and dark matter and the corresponding equations of state (EoS). Section three briefly explains how these EoS are used to calculate the properties of the neutron stars (NS). This is followed by section four, where the NS properties are calculated and the effects of neutron decay on those properties are studied. Section five summarises the main conclusions of this study.

\section{Neutron star model and Equation of state (EoS)}
Neutron stars are often considered to be laboratories for testing the physics at ultra-high densities because they contain matter in extreme conditions, which cannot be prepared in a laboratory. Neutron stars cover a wide range of energy densities, possibly exceeding six times nuclear matter density at the core. At such high densities there is no consensus concerning the state of matter, for example whether it is still hadronic or has made a transition to deconfined quark matter~\cite{doi:10.1143/JPSJ.58.3555,Bombaci_2004,2012_a,Masuda:2012ed,doi:10.1063/1.4909561,2016_a,PhysRevC.58.1804,PhysRevLett.79.1603,weber2007neutron,Whittenbury:2015ziz,Kapusta:2021ney,Li:2018ayl,Tanimoto:2019tsl,2019_fri,Husain_2021,2001_lattimer,2020_latti,10.1143/PTP.108.703,2017_xyz,Motta:2022nlj,2022_xxyz}. Here we consider two EoSs which are based on the quark meson coupling model, one involving nucleons only and the other including hyperons at the higher energy densities in the core. The details of the QMC model can be found in Refs.~\cite{Guichon:2018uew,Stone:2016qmi,Guichon:1995ue,RIKOVSKASTONE2007341,Martinez:2020ctv,Martinez:2018xep,Saito:1998eu,TSUSHIMA1998691}. For our purposes it is sufficient that these EoS produce NS with maximum mass, tidal deformability and radii consistent~\cite{Motta:2019tjc,Motta:2020xsg} with current constraints from gravitational 
wave~\cite{LIGOScientific:2017vwq,LIGOScientific:2018cki} and satellite observations~\cite{Riley:2021pd}.

Assuming that the matter in the NS is hadronic, it is mostly made of neutrons and if the neutrons decay into dark matter then there should be enough dark matter particles inside the neutron star to make some changes in its properties~\cite{CIARCELLUTI201119,Sandin_2009,PhysRevD.77.043515,Mukhopadhyay_2017,Leung:2011zz,Ellis_2018,Kouvaris_2012,2021_w,2000NuPhB.564..185M,2018_sa,berryman2022neutron,2019_reddy,2013_red,deLavallaz:2010wp,Sen:2021wev,PhysRevD.105.023001}. Therefore, we may expect to test this model for neutron decay using the properties of neutron stars.

\subsection{Neutron decay model}
Following Strumia~\cite{Strumia_2022}, we consider the possible decay of the neutron as
\begin{equation}
    n \rightarrow \chi\chi\chi,
\end{equation}
where $\chi$ is a dark matter fermion with baryon number 1/3. The mass of dark matter particle must be extremely close to $m_n$/3, in order to ensure that nuclei remain stable. For this decay to be kinematically allowed the mass of the dark matter particle must be $m_\chi < m_n/3 \approx$ 313.19 MeV, while for the nuclear decay to be forbidden one must have $m_\chi$ $>$ ($m_n - E_{Be}$)/3 $\approx$ 312.63 MeV, because the strongest bound comes from $^8$Be. Stability of the proton requires that $m_\chi  > (m_p - m_e)$/3 = 312.59 MeV, and for $m_\chi < (m_p + m_e
)$/3 $\approx$ 312.93 MeV the decay of a Hydrogen atom, H $\rightarrow \chi\chi\chi\nu_e$, would be allowed. While it is difficult to argue that such fine tuning yields a natural explanation of the neutron lifetime puzzle, it is important to explore every possible test of the idea.

\subsection{Equation of state}
The decay of neutrons into dark matter will change the chemical composition of neutron star. The conditions of $\beta$-equilibrium inside the neutron star require that the following equations hold for the 
nucleon-only EoS,
\begin{equation}
 n_p = n_e + n_\mu \, , \qquad
 \mu_n = \mu_p + \mu_e, \qquad
 \mu_\mu = \mu_e, \qquad
  \mu_\chi =  \mu_n/3  \qquad \, . 
\end{equation}
For simplicity we choose $m_\chi$ = $m_n$/3 in all numerical calculations. For our second EoS for nuclear matter, we allow hyperons to appear as the chemical potential increases. With hyperons included in the EoS we must have
\begin{eqnarray}
 n_p = n_e + n_\mu + n_{\Xi^-}\, , \qquad
 \mu_n = \mu_p + \mu_e, \qquad
 \mu_\mu = \mu_e, \qquad
  \mu_\chi =  \mu_n/3,  \qquad \cr
  \mu_n = \mu_{\Lambda} = \mu_{\Xi^0} \qquad \, ,
\end{eqnarray}
and $\mu_{\Xi^-} = \mu_n + \mu_e$. We note that the repulsion experienced by the $\Sigma$ baryons in the QMC model~\cite{Guichon:2008zz} means that they do not appear at any density of interest and so have not been shown explicitly. The hyperons begin to appear when the density exceeds 3 times nuclear matter density, with the $\Lambda$ appearing first, followed by the $\Xi^-$ and only at the highest central densities the $\Xi^0$. The energy 
density~\cite{Motta:2019tjc,Husain:2022bxl} is calculated using the Hartree-Fock method. The Hartree term is as follows 
\begin{eqnarray}
    \epsilon_H = \frac{1}{2}m^2_\sigma\sigma^2 + \frac{1}{2}m^2_\omega\omega^2 + 
    \frac{1}{2}m^2_\rho\rho^2 + \frac{1}{\pi^2} \int^{k_F^n}_0 k^2 \sqrt{k^2 + M_N^*(\sigma)^2} dk + \cr
    \frac{1}{\pi^2} \int^{k_F^p}_0 k^2 \sqrt{k^2 + M_N^*(\sigma)^2} dk  +  \frac{1}{\pi^2}\int^{k_F^e}_0 k^2\sqrt{k^2 + m_e^2} dk \cr + \frac{1}{\pi^2} \int^{k_F^\mu}_0 k^2 \sqrt{k^2 + m_\mu^2} dk + 
    \frac{1}{\pi^2}\int^{k_F^\chi}_0 k^2\sqrt{k^2 + m_\chi^2} dk 
    \, ,
\end{eqnarray}
where $M_N^*$ is the effective mass of the nucleon given by
\begin{equation}
M_N^*(\sigma) = M_N -  g_\sigma \sigma + 
\frac{d (g_\sigma \sigma)^2}{2} \, . 
\end{equation}
Here $M_N$ is the mass of the nucleon, $g_\sigma$ is the scalar ($\sigma$) coupling to the nucleon in free space and $d$ is the scalar polarisability, which arises because of the change in nucleon structure caused by the strong scalar mean field in dense nuclear matter~\cite{Guichon:1987jp,Guichon:1995ue}. The expressions for the Fock terms (including those for the pion), which are more complicated, can be found in Ref.~\cite{Motta:2019tjc}. The pressure is calculated using 
\begin{equation}
    P = \int \mu_f n_f - \epsilon.
\end{equation}
 and the baryon number is given by 
\begin{equation}
    B_i = 4\pi\int_0^R \frac{r^2 n_i(r)}{[1 - \frac{2M(r)}{r}]^{1/2}} 
    dr \, ,
\end{equation}
where $i$ identifies the different baryons, such as neutron, proton and hyperons, while the dark matter baryon number involves the number density divided by a factor of three.

\section{Structural equations for the neutron star}
In order to avoid additional unknown parameters, we make the working hypothesis that the nuclear matter does not interact with the dark matter. Hence, the two fluid TOV~\cite{PhysRev.55.374,Tolman169} equations must be used to calculate the distribution of matter in the star. The following two fluid coupled TOV equations are integrated from the centre of the neutron star to the surface. 
\begin{equation}
\frac{dP_{nucl}}{dr} = - \frac{[\epsilon_{nucl}(r)+P_{nucl}(r)][4\pi r^3(P_{nucl}(r) +  P_{DM}(r))+m(r)]}{r^2(1-\frac{2m(r)}{r})},
\end{equation}
\begin{equation}
m_{nucl}(r) = 4\pi\int_{0}^{r}dr.r^2 \epsilon_{nucl}(r) \, ,
\end{equation}
\begin{equation}
\frac{dP_{DM}}{dr} = - \frac{[\epsilon_{DM}(r)+P_{DM}(r)][4\pi r^3(P_{nucl}(r) +  P_{DM}(r))+m(r)]}{r^2(1-\frac{2m(r)}{r})} \, ,
\end{equation}
\begin{equation}
m_{DM}(r) = 4\pi\int_{0}^{r}dr.r^2 \epsilon_{DM}(r) \, ,
\end{equation}
\begin{equation}
m(r) = m_{nucl}(r) + m_{DM}(r) \, .
\end{equation}
For the purpose of calculating the tidal deformability, the method provided in Refs.~\cite{Hinderer_2008,Hinderer_2010} is considered. 

\section{Effects of neutron decay}
\begin{figure}[ht]
\centering 
\includegraphics[width=1\textwidth]{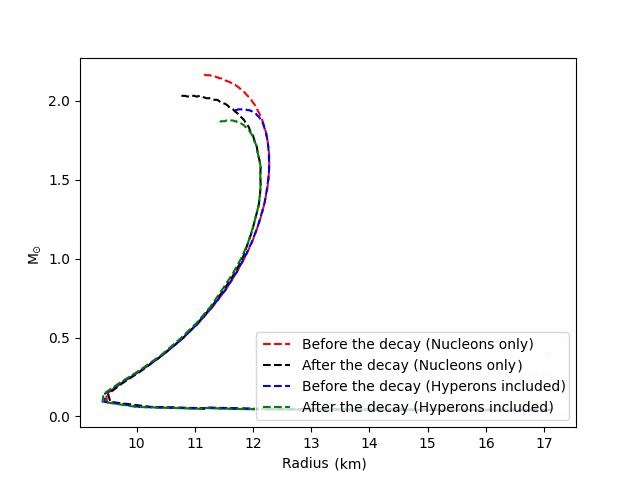}
\caption{Total mass versus radius of a NS, at T = 0 $^\circ K$, before and after the neutrons decay.}
\label{fig1}
\end{figure}

The effects of neutron decay on the neutron star properties are presented in this section. Figure~\ref{fig1} shows that as the neutrons decay into dark matter the maximum mass of the neutron star stays close to 2M$_\odot$, whether we use the nucleon-only EoS or hyperons-included EoS. The presence of hyperons in the core makes the equation of state softer because the high momentum neutrons are replaced by low momentum hyperons and as the energy density increases the neutron star cannot sustain itself against gravity and collapses into a black hole. This is why the nucleon-only equation of state (EoS) gives a neutron star of greater maximum mass compared to that with hyperons included. This is also why the decay of neutrons into dark matter makes the EoS even softer, whether nucleons only or hyperons are included in the EoS. The repulsive interaction among the dark matter particles is not sufficient to increase the neutron star mass to where it was without the neutron decay. Hence after the neutrons decay the neutron stars have lower mass. Observations have put significant constraints on neutron star properties so that for an EoS to be considered realistic at higher energy densities it must respect those constraints. In particular, the maximum mass of the neutron star should be at least 2 solar masses, the radius has to be in the range of 10-14 kms \cite{Abbott_2017} and tidal deformability must be in the range 75-580. Therefore, in agreement with Strumia~\cite{Strumia_2022}, we find that without any need for additional self repulsion between the dark matter particles, the NS can maintain a maximum mass in the region of  2M$_\odot$, which is our maximum mass constraint. 

\begin{figure}[ht]
\centering 
\includegraphics[width=1\textwidth]{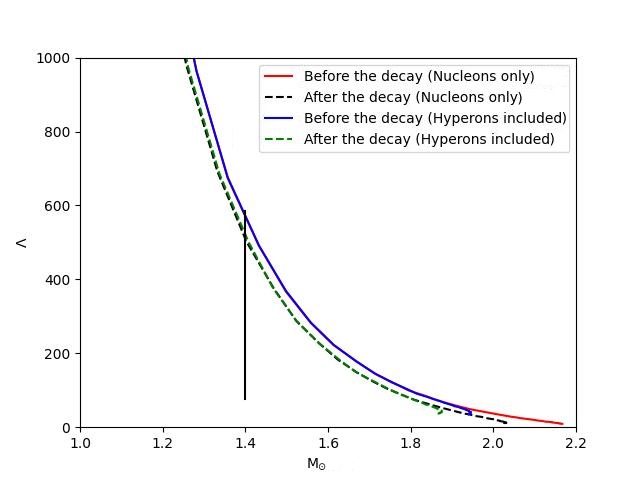}
\caption{Total mass (M$_{\odot}$) versus tidal deformability of neutron star before and after the decay of neutrons.}
\label{fig4}
\end{figure}

Figure ~\ref{fig4} tests the hypothesis of neutron decay against the tidal deformability constraint~\cite{Abbott_2017,Abbott_2019,Bramante_2018_31}, namely that a neutron star of mass 1.4 M$_\odot$ should have a tidal deformability in the range $70 \le \Lambda \le 580$ . For both EoS the value of tidal deformability of neutron star after the decay decreases slightly, clearly satisfying this tidal deformability constraint.
%

\begin{figure}[ht]
\centering 
\includegraphics[width=1\textwidth]{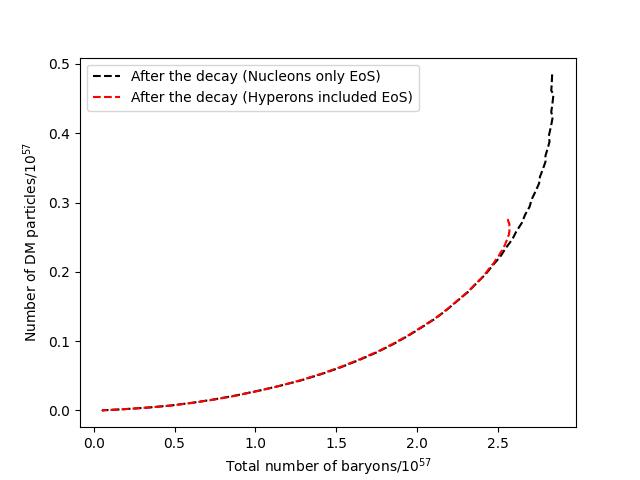}
\caption{Total number of baryons versus total number of dark matter particles. Each dark matter particle has baryon number $\frac{1}{3}$.}
\label{fig6}
\end{figure}

Figure~\ref{fig6} shows that in the lighter neutron stars the fraction of the matter converted to dark matter is quite small, although it increases significantly in heavier neutron stars. In maximum mass neutron stars produced by both hadronic EoS, we find around 4\% of the total mass contributed by the dark matter. The dark matter particles contribute about 12\% of the total number of particles which is large enough to generate enough repulsion to sustain the neutron star against gravity and satisfy the constraints on maximum mass and tidal deformability of the neutron star.
%

\begin{figure}[ht]
\centering 
\includegraphics[width=1\textwidth]{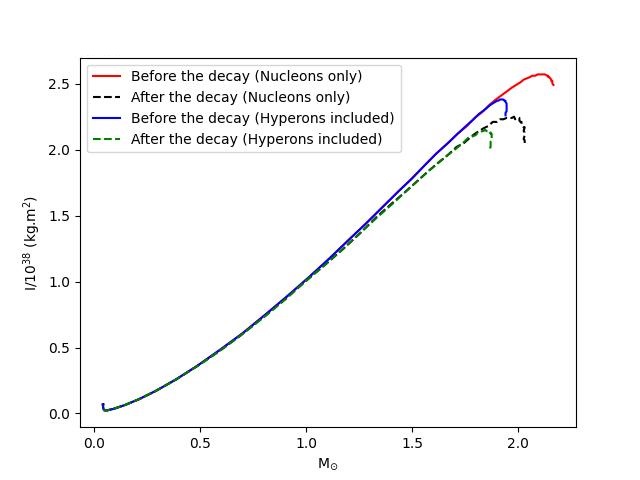}
\caption{Total mass versus moment of inertia of neutron star before and after the neutrons decay.}
\label{fig5}
\end{figure}

Figure~\ref{fig5} illustrates the moment of inertia for both hadronic EoS. Clearly the neutron decay leads to a decrease in the moment of inertia of the NS and therefore the spin of neutron star must increase after the decay. This is expected since there is significant reduction in the mass and radius of the neutron star after the decay, as shown in Figure ~\ref{fig1}. The enhancement of the spin and the reduction in rotational period could play a vital role in detecting a signal of neutron decay inside the neutron star. 
\begin{figure}[ht]
\centering 
\includegraphics[width=1\textwidth]{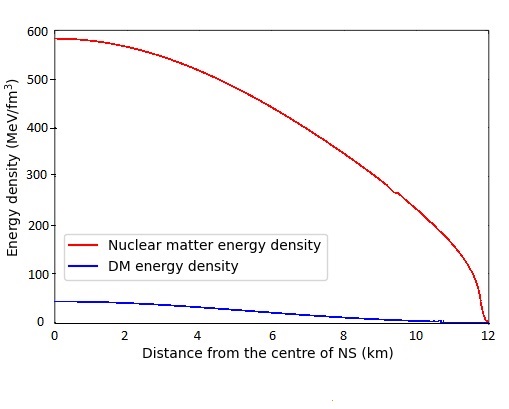}
\caption{Distribution of nuclear matter and dark matter inside the neutron star at T = 0 $^\circ K$ after the neutrons decay for nucleons only EoS.}
\label{fig7}
\end{figure}
%

Figure~\ref{fig7} displays the distribution of energy density of the nuclear matter and dark matter from the centre of the neutron star to the surface. Although we only show the nucleon-only result, the hyperons-included case is essentially identical. Both selected EoS, be it nucleon-only or hyperons-included, show that the degenerate dark matter remains inside and does not envelop the the neutron star. Most of the dark matter is trapped inside the inner core of the NS.  

As the neutron decay takes place the total energy and the total baryon number inside the NS must remain conserved. The values given in Table~\ref{Table:1} correspond to a 1.8 M$_\odot$ NS and show that for both EoS, if the total baryon number is kept fixed, then after the decay the mass of the neutron star is lighter than before, by approximately 0.007M$_\odot$. Since neutron stars are very compact objects and have an escape velocity comparable to the speed of light, there is no mechanism that can give particles enough velocity to escape the star following the decay. 

The solution to this problem is that the dark matter will not be completely degenerate but have effectively a finite temperature. The hot dark matter inside the neutron star then contains the energy equivalent to  the difference of masses of neutron star before and after the decay, that is 0.007M$_\odot$. We calculated the NS properties with hot dark matter having heat energy equivalent to 0.007M$_\odot$ and they are summarised in Table ~\ref{Table:2}. The dark matter inside the neutron must have a temperature of approximately 6.5 MeV and 6 MeV for nucleon-only and hyperons-included EoS, respectively. As there is no significant change in the radius of the NS associated with dark matter heating, the moment of inertia of the star at 6 MeV or 6.5 MeV remains almost the same as the moment of inertia at 0$^\circ K$. However, because of the decay the change in radius of the neutron star is significant and the total change in moment of inertia is approximately 5.75\%. As a result the neutron star must spin up during the decay. 
\begin{center}
	\begin{table}
		\begin{tabular}{|c|c|c|c|c|}
			\hline
		& NS Mass & T (MeV) &  R (Km) & I (kg.m$^2$)   \\
		\hline
		
		1 &	1.8 M$_\odot$ (Nucleon only EoS before decay) & 0  & 12.25 & 2.25 $\times$ 10$^{38}$ \\
		\hline
		2 &	1.793644 M$_\odot$ (Nucleon only EoS + DM after decay) & 0  & 11.96 & 2.12 $\times$ 10$^{38}$    \\
		\hline
		3 &	1.8 M$_\odot$ (Hyperons included EoS  before decay) & 0  & 12.20 & 2.24 $\times$ 10$^{38}$    \\
		\hline
		4 &	1.793577 M$_\odot$ (Hyperons included EoS + DM after decay) & 0  & 11.96 & 2.11 $\times$ 10$^{38}$    \\
		\hline
		\end{tabular}
		\caption{Properties of a 1.8M$_\odot$ neutron star with total baryon number 2.4445 $\times 10^{57}$ at temperature 0 MeV,   before and after the neutrons decay. Here, T stands for the temperature and M represents the NS mass, while R and I denote the radius and the moment of inertia of the neutron star.}
		\label{Table:1}
	\end{table}
\end{center}
\begin{center}
	\begin{table}
		\begin{tabular}{|c|c|c|c|}
			\hline
		 M$_\odot$ & T (MeV)  of DM &  R (Km) & I (kg.m$^2$)   \\
		\hline
		
			1.8 M$_\odot$ (Nucleon only EoS before decay) & 0  & 12.25 & 2.25 $\times$ 10$^{38}$ \\
		\hline
			1.793644 M$_\odot$ (Nucleon only EoS + DM after decay) & 6.5  & 11.96 & 2.124 $\times$ 10$^{38}$    \\
		\hline
			1.8 M$_\odot$ (Hyperons included EoS  before decay) & 0  & 12.20 & 2.24 $\times$ 10$^{38}$    \\
		\hline
			1.793577 M$_\odot$ (Hyperons included EoS + DM after decay) & 6  & 11.96 & 2.115 $\times$ 10$^{38}$    \\
		\hline
		\end{tabular}
		\caption{Properties of a 1.8M$_\odot$ neutron star with total baryon number 2.4445 $\times 10^{57}$, before and after the neutrons decay. Here, T stands for the temperature of the dark matter and M represents the NS mass, while R and I denotes the radius and the moment of inertia of the neutron star.}
		\label{Table:2}
	\end{table}
\end{center}

\section{Conclusion}
From Fig.~\ref{fig1} it is evident that the decay of a neutron into three dark matter particles is consistent with the neutron star maximum mass constraint, without any need for dark matter self-repulsion. Whether it is  hyperons or nucleons in the core, the dark matter can sustain the neutron star against gravitational collapse and give a NS with a maximum mass of approximately 2 $M_\odot$. Figure~\ref{fig4} shows that the neutron decay into dark matter is also consistent with the tidal deformability constraint.

As illustrated by Fig.~\ref{fig6} the presence of dark matter increases rapidly in heavier neutron stars, suggesting that we should investigate heavier neutron stars to study this mode of neutron decay.  After the decay the moment of inertia of the neutron star is reduced, as illustrated in Fig.~\ref{fig5}, which requires that after the decay the NS should spin up. From Fig.~\ref{fig7} we see that the dark matter remains inside the NS, with most of the dark matter inside the inner core.

From the conservation of total baryon number and total energy, Tables ~\ref{Table:1} and ~\ref{Table:2} indicate that for the decay to take place the dark matter inside the neutron must heat up by around 6 MeV. Therefore, a tiny amount of heat will be radiated in a few seconds by the Urca process while the DM will take considerably longer time to cool. The DM matter cooling will primarily take place by DM-DM collision creating neutrons which are only partially Pauli blocked due rapid nuclear matter cooling because the cooling of nuclear matter will be much faster than the decay rate. Although the mechanism of DM cooling will be slower than the rate of neutron decay into dark matter, yet it will cool down the neutron star eventually. 
Although the heating of dark matter does not cause a significant change in radius or moment of inertia compared to the neutron star properties at 0$^\circ K$, the total change in radius and moment of inertia before and after the decay is significant. Therefore, if neutrons decay into dark matter in this way, the NS must spin-up. Whether or not such a spin-up can be observed is a challenge for future work. 

\section*{Acknowledgements}
This work was supported by the University of Adelaide International Scholarship and by the Australian Research Council through grants CE200100008 and DP180100497.

\section{References}
\bibliography{main.bib}

\begin{thebibliography}{10}

\bibitem{RevModPhys.83.1173}
F.~E. Wietfeldt and G.~L. Greene, ``Colloquium: The neutron lifetime,'' {\em
  Rev. Mod. Phys.}, vol.~83, pp.~1173--1192, Nov 2011.

\bibitem{Otono:2016fsv}
H.~Otono, ``{LiNA \textendash{} Lifetime of neutron apparatus with time
  projection chamber and solenoid coil},'' {\em Nucl. Instrum. Meth. A},
  vol.~845, pp.~278--280, 2017.

\bibitem{PhysRevLett.111.222501}
A.~T. Yue, M.~S. Dewey, D.~M. Gilliam, G.~L. Greene, A.~B. Laptev, J.~S. Nico,
  W.~M. Snow, and F.~E. Wietfeldt, ``Improved determination of the neutron
  lifetime,'' {\em Phys. Rev. Lett.}, vol.~111, p.~222501, Nov 2013.

\bibitem{Byrne_1996}
J.~Byrne, P.~G. Dawber, C.~G. Habeck, S.~J. Smidt, J.~A. Spain, and A.~P.
  Williams, ``A revised value for the neutron lifetime measured using a penning
  trap,'' {\em Europhysics Letters ({EPL})}, vol.~33, pp.~187--192, jan 1996.

\bibitem{PhysRevC.71.055502}
J.~S. Nico, M.~S. Dewey, D.~M. Gilliam, F.~E. Wietfeldt, X.~Fei, W.~M. Snow,
  G.~L. Greene, J.~Pauwels, R.~Eykens, A.~Lamberty, J.~V. Gestel, and R.~D.
  Scott, ``Measurement of the neutron lifetime by counting trapped protons in a
  cold neutron beam,'' {\em Phys. Rev. C}, vol.~71, p.~055502, May 2005.

\bibitem{SEREBROV200572}
A.~Serebrov, V.~Varlamov, A.~Kharitonov, A.~Fomin, Y.~Pokotilovski,
  P.~Geltenbort, J.~Butterworth, I.~Krasnoschekova, M.~Lasakov, R.~Tal'daev,
  A.~Vassiljev, and O.~Zherebtsov, ``Measurement of the neutron lifetime using
  a gravitational trap and a low-temperature fomblin coating,'' {\em Physics
  Letters B}, vol.~605, no.~1, pp.~72--78, 2005.

\bibitem{PICHLMAIER2010221}
A.~Pichlmaier, V.~Varlamov, K.~Schreckenbach, and P.~Geltenbort, ``Neutron
  lifetime measurement with the ucn trap-in-trap mambo ii,'' {\em Physics
  Letters B}, vol.~693, no.~3, pp.~221--226, 2010.

\bibitem{PhysRevC.85.065503}
A.~Steyerl, J.~M. Pendlebury, C.~Kaufman, S.~S. Malik, and A.~M. Desai,
  ``Quasielastic scattering in the interaction of ultracold neutrons with a
  liquid wall and application in a reanalysis of the mambo i neutron-lifetime
  experiment,'' {\em Phys. Rev. C}, vol.~85, p.~065503, Jun 2012.

\bibitem{ARZUMANOV201579}
S.~Arzumanov, L.~Bondarenko, S.~Chernyavsky, P.~Geltenbort, V.~Morozov,
  V.~Nesvizhevsky, Y.~Panin, and A.~Strepetov, ``A measurement of the neutron
  lifetime using the method of storage of ultracold neutrons and detection of
  inelastically up-scattered neutrons,'' {\em Physics Letters B}, vol.~745,
  pp.~79--89, 2015.

\bibitem{UCNt:2021pcg}
F.~M. Gonzalez {\em et~al.}, ``{Improved Neutron Lifetime Measurement with
  UCN\ensuremath{\tau}},'' {\em Phys. Rev. Lett.}, vol.~127, no.~16, p.~162501,
  2021.

\bibitem{scienceaan8895}
R.~W. Pattie, N.~B. Callahan, C.~Cude-Woods, E.~R. Adamek, L.~J. Broussard,
  S.~M. Clayton, S.~A. Currie, E.~B. Dees, X.~Ding, E.~M. Engel, D.~E. Fellers,
  W.~Fox, P.~Geltenbort, K.~P. Hickerson, M.~A. Hoffbauer, A.~T. Holley,
  A.~Komives, C.-Y. Liu, S.~W.~T. MacDonald, M.~Makela, C.~L. Morris, J.~D.
  Ortiz, J.~Ramsey, D.~J. Salvat, A.~Saunders, S.~J. Seestrom, E.~I. Sharapov,
  S.~K. Sjue, Z.~Tang, J.~Vanderwerp, B.~Vogelaar, P.~L. Walstrom, Z.~Wang,
  W.~Wei, H.~L. Weaver, J.~W. Wexler, T.~L. Womack, A.~R. Young, and B.~A.
  Zeck, ``Measurement of the neutron lifetime using a magneto-gravitational
  trap and in situ detection,'' {\em Science}, vol.~360, no.~6389,
  pp.~627--632, 2018.

\bibitem{PhysRevC.97.055503}
A.~P. Serebrov, E.~A. Kolomensky, A.~K. Fomin, I.~A. Krasnoshchekova, A.~V.
  Vassiljev, D.~M. Prudnikov, I.~V. Shoka, A.~V. Chechkin, M.~E. Chaikovskiy,
  V.~E. Varlamov, S.~N. Ivanov, A.~N. Pirozhkov, P.~Geltenbort, O.~Zimmer,
  T.~Jenke, M.~Van~der Grinten, and M.~Tucker, ``Neutron lifetime measurements
  with a large gravitational trap for ultracold neutrons,'' {\em Phys. Rev. C},
  vol.~97, p.~055503, May 2018.

\bibitem{PhysRevLett.120.191801}
B.~Fornal and B.~Grinstein, ``Dark matter interpretation of the neutron decay
  anomaly,'' {\em Phys. Rev. Lett.}, vol.~120, p.~191801, May 2018.

\bibitem{doi:10.1142/S0217732320300190}
B.~Fornal and B.~Grinstein, ``Neutron’s dark secret,'' {\em Modern Physics
  Letters A}, vol.~35, no.~31, p.~2030019, 2020.

\bibitem{Rajendran_2021}
S.~Rajendran and H.~Ramani, ``Composite solution to the neutron lifetime
  anomaly,'' {\em Physical Review D}, vol.~103, feb 2021.

\bibitem{Serebrov:2018mva}
A.~P. Serebrov, R.~M. Samoilov, I.~A. Mitropolsky, and A.~M. Gagarsky,
  ``{Neutron lifetime, dark matter and search for sterile neutrino},'' 2 2018.

\bibitem{2018_tanga}
Z.~Tang, M.~Blatnik, L.~Broussard, J.~Choi, S.~Clayton, C.~Cude-Woods,
  S.~Currie, D.~Fellers, E.~Fries, P.~Geltenbort, F.~Gonzalez, K.~Hickerson,
  T.~Ito, C.-Y. Liu, S.~MacDonald, M.~Makela, C.~Morris, C.~O’Shaughnessy,
  R.~Pattie, B.~Plaster, D.~Salvat, A.~Saunders, Z.~Wang, A.~Young, and
  B.~Zeck, ``Search for the neutron decay n $\rightarrow \chi$+$\gamma$ , where
  $\chi$ is a dark matter particle,'' {\em Physical Review Letters}, vol.~121,
  Jul 2018.

\bibitem{Agostini_2015}
M.~Agostini, S.~Appel, G.~Bellini, J.~Benziger, D.~Bick, G.~Bonfini, D.~Bravo,
  B.~Caccianiga, F.~Calaprice, A.~Caminata, P.~Cavalcante, A.~Chepurnov,
  D.~D'Angelo, S.~Davini, A.~Derbin, L.~D. Noto, I.~Drachnev, A.~Empl,
  A.~Etenko, K.~Fomenko, D.~Franco, F.~Gabriele, C.~Galbiati, C.~Ghiano,
  M.~Giammarchi, M.~Goeger-Neff, A.~Goretti, M.~Gromov, C.~Hagner,
  E.~Hungerford, A.~Ianni, A.~Ianni, K.~Jedrzejczak, M.~Kaiser, V.~Kobychev,
  D.~Korablev, G.~Korga, D.~Kryn, M.~Laubenstein, B.~Lehnert, E.~Litvinovich,
  F.~Lombardi, P.~Lombardi, L.~Ludhova, G.~Lukyanchenko, I.~Machulin,
  S.~Manecki, W.~Maneschg, S.~Marcocci, E.~Meroni, M.~Meyer, L.~Miramonti,
  M.~Misiaszek, M.~Montuschi, P.~Mosteiro, V.~Muratova, B.~Neumair,
  L.~Oberauer, M.~Obolensky, F.~Ortica, K.~Otis, M.~Pallavicini, L.~Papp,
  L.~Perasso, A.~Pocar, G.~Ranucci, A.~Razeto, A.~Re, A.~Romani, R.~Roncin,
  N.~Rossi, S.~Schönert, D.~Semenov, H.~Simgen, M.~Skorokhvatov, O.~Smirnov,
  A.~Sotnikov, S.~Sukhotin, Y.~Suvorov, R.~Tartaglia, G.~Testera, J.~Thurn,
  M.~Toropova, E.~Unzhakov, A.~Vishneva, R.~Vogelaar, F.~von Feilitzsch,
  H.~Wang, S.~Weinz, J.~Winter, M.~Wojcik, M.~Wurm, Z.~Yokley, O.~Zaimidoroga,
  S.~Zavatarelli, K.~Zuber, and G.~Z. and, ``Test of electric charge
  conservation with borexino,'' {\em Physical Review Letters}, vol.~115, dec
  2015.

\bibitem{Motta:2018bil}
T.~F. Motta, P.~A.~M. Guichon, and A.~W. Thomas, ``{Neutron to Dark Matter
  Decay in Neutron Stars},'' {\em Int. J. Mod. Phys. A}, vol.~33, no.~31,
  p.~1844020, 2018.

\bibitem{Motta:2018rxp}
T.~F. Motta, P.~A.~M. Guichon, and A.~W. Thomas, ``{Implications of Neutron
  Star Properties for the Existence of Light Dark Matter},'' {\em J. Phys. G},
  vol.~45, no.~5, p.~05LT01, 2018.

\bibitem{PhysRevLett.121.061801}
G.~Baym, D.~H. Beck, P.~Geltenbort, and J.~Shelton, ``Testing dark decays of
  baryons in neutron stars,'' {\em Phys. Rev. Lett.}, vol.~121, p.~061801, Aug
  2018.

\bibitem{2018_sa}
D.~McKeen, A.~E. Nelson, S.~Reddy, and D.~Zhou, ``Neutron stars exclude light
  dark baryons,'' {\em Physical Review Letters}, vol.~121, Aug 2018.

\bibitem{Strumia_2022}
A.~Strumia, ``Dark matter interpretation of the neutron decay anomaly,'' {\em
  Journal of High Energy Physics}, vol.~2022, feb 2022.

\bibitem{doi:10.1143/JPSJ.58.3555}
H.~Terazawa, ``Super-hypernuclei in the quark-shell model,'' {\em Journal of
  the Physical Society of Japan}, vol.~58, no.~10, pp.~3555--3563, 1989.

\bibitem{Bombaci_2004}
I.~Bombaci, I.~Parenti, and I.~Vidana, ``Quark deconfinement and implications
  for the radius and the limiting mass of compact stars,'' {\em The
  Astrophysical Journal}, vol.~614, pp.~314--325, oct 2004.

\bibitem{2012_a}
I.~Bednarek, P.~Haensel, J.~L. Zdunik, M.~Bejger, and R.~Mańka, ``Hyperons in
  neutron-star cores and a 2m pulsar,'' {\em Astronomy \& Astrophysics},
  vol.~543, p.~A157, Jul 2012.

\bibitem{Masuda:2012ed}
K.~Masuda, T.~Hatsuda, and T.~Takatsuka, ``{Hadron\textendash{}quark crossover
  and massive hybrid stars},'' {\em PTEP}, vol.~2013, no.~7, p.~073D01, 2013.

\bibitem{doi:10.1063/1.4909561}
I.~Vidaña, ``Hyperons and neutron stars,'' {\em AIP Conference Proceedings},
  vol.~1645, no.~1, pp.~79--85, 2015.

\bibitem{2016_a}
M.~Oertel, F.~Gulminelli, C.~Providência, and A.~R. Raduta, ``Hyperons in
  neutron stars and supernova cores,'' {\em The European Physical Journal A},
  vol.~52, Mar 2016.

\bibitem{PhysRevC.58.1804}
A.~Akmal, V.~R. Pandharipande, and D.~G. Ravenhall, ``Equation of state of
  nucleon matter and neutron star structure,'' {\em Phys. Rev. C}, vol.~58,
  pp.~1804--1828, Sep 1998.

\bibitem{PhysRevLett.79.1603}
N.~K. Glendenning, S.~Pei, and F.~Weber, ``Signal of quark deconfinement in the
  timing structure of pulsar spin-down,'' {\em Phys. Rev. Lett.}, vol.~79,
  pp.~1603--1606, Sep 1997.

\bibitem{weber2007neutron}
F.~Weber, R.~Negreiros, and P.~Rosenfield, ``Neutron star interiors and the
  equation of state of superdense matter,'' 2007.

\bibitem{Whittenbury:2015ziz}
D.~L. Whittenbury, H.~H. Matevosyan, and A.~W. Thomas, ``{Hybrid stars using
  the quark-meson coupling and proper-time Nambu\textendash{}Jona-Lasinio
  models},'' {\em Phys. Rev. C}, vol.~93, no.~3, p.~035807, 2016.

\bibitem{Kapusta:2021ney}
J.~I. Kapusta and T.~Welle, ``{Neutron stars with a crossover equation of
  state},'' {\em Phys. Rev. C}, vol.~104, no.~1, p.~L012801, 2021.

\bibitem{Li:2018ayl}
C.-M. Li, Y.~Yan, J.-J. Geng, Y.-F. Huang, and H.-S. Zong, ``{Constraints on
  the hybrid equation of state with a crossover hadron-quark phase transition
  in the light of GW170817},'' {\em Phys. Rev. D}, vol.~98, no.~8, p.~083013,
  2018.

\bibitem{Tanimoto:2019tsl}
T.~Tanimoto, W.~Bentz, and I.~C. Clo\"et, ``{Massive Neutron Stars with a Color
  Superconducting Quark Matter Core},'' {\em Phys. Rev. C}, vol.~101, no.~5,
  p.~055204, 2020.

\bibitem{2019_fri}
W.~M. Spinella and F.~Weber, ``Hyperonic neutron star matter in light of
  gw170817,'' {\em Astronomische Nachrichten}, vol.~340, p.~145–150, Jan
  2019.

\bibitem{Husain_2021}
W.~Husain and A.~W. Thomas, ``{Hybrid Stars with Hyperons and Strange Quark
  Matter},'' {\em AIP Conf. Proc.}, vol.~2319, no.~1, p.~080001, 2021.

\bibitem{2001_lattimer}
J.~M. Lattimer and M.~Prakash, ``Neutron star structure and the equation of
  state,'' {\em The Astrophysical Journal}, vol.~550, p.~426–442, Mar 2001.

\bibitem{2020_latti}
T.~Zhao and J.~M. Lattimer, ``Quarkyonic matter equation of state in
  beta-equilibrium,'' {\em Physical Review D}, vol.~102, Jul 2020.

\bibitem{10.1143/PTP.108.703}
S.~Nishizaki, Y.~Yamamoto, and T.~Takatsuka, ``{Hyperon-Mixed Neutron Star
  Matter and Neutron Stars*)},'' {\em Progress of Theoretical Physics},
  vol.~108, pp.~703--718, 10 2002.

\bibitem{2017_xyz}
Y.~Yamamoto, H.~Togashi, T.~Tamagawa, T.~Furumoto, N.~Yasutake, and T.~A.
  Rijken, ``Neutron-star radii based on realistic nuclear interactions,'' {\em
  Physical Review C}, vol.~96, Dec 2017.

\bibitem{Motta:2022nlj}
T.~F. Motta and A.~W. Thomas, ``{The role of baryon structure in neutron
  stars},'' {\em Mod. Phys. Lett. A}, vol.~37, no.~01, p.~2230001, 2022.

\bibitem{2022_xxyz}
Y.~Yamamoto, N.~Yasutake, and T.~A. Rijken, ``Quark-quark interaction and quark
  matter in neutron stars,'' {\em Physical Review C}, vol.~105, Jan 2022.

\bibitem{Guichon:2018uew}
P.~A.~M. Guichon, J.~R. Stone, and A.~W. Thomas,
  ``{Quark\textendash{}Meson-Coupling (QMC) model for finite nuclei, nuclear
  matter and beyond},'' {\em Prog. Part. Nucl. Phys.}, vol.~100, pp.~262--297,
  2018.

\bibitem{Stone:2016qmi}
J.~R. Stone, P.~A.~M. Guichon, P.~G. Reinhard, and A.~W. Thomas, ``{Finite
  Nuclei in the Quark-Meson Coupling Model},'' {\em Phys. Rev. Lett.},
  vol.~116, no.~9, p.~092501, 2016.

\bibitem{Guichon:1995ue}
P.~A.~M. Guichon, K.~Saito, E.~N. Rodionov, and A.~W. Thomas, ``{The Role of
  nucleon structure in finite nuclei},'' {\em Nucl. Phys. A}, vol.~601,
  pp.~349--379, 1996.

\bibitem{RIKOVSKASTONE2007341}
J.~{Rikovska Stone}, P.~Guichon, H.~Matevosyan, and A.~Thomas, ``Cold uniform
  matter and neutron stars in the quark–meson-coupling model,'' {\em Nuclear
  Physics A}, vol.~792, no.~3, pp.~341--369, 2007.

\bibitem{Martinez:2020ctv}
K.~M.~L. Martinez, A.~W. Thomas, P.~A.~M. Guichon, and J.~R. Stone, ``{Tensor
  and pairing interactions within the quark-meson coupling energy-density
  functional},'' {\em Phys. Rev. C}, vol.~102, no.~3, p.~034304, 2020.

\bibitem{Martinez:2018xep}
K.~M.~L. Martinez, A.~W. Thomas, J.~R. Stone, and P.~A.~M. Guichon,
  ``{Parameter optimization for the latest quark-meson coupling energy-density
  functional},'' {\em Phys. Rev. C}, vol.~100, no.~2, p.~024333, 2019.

\bibitem{Saito:1998eu}
K.~Saito, ``{The quark - meson coupling model},'' in {\em {14th International
  Baldin Seminar on High Energy Physics Problems}: {Relativistic Nuclear
  Physics and Quantum Chromodynamics}}, 8 1998.

\bibitem{TSUSHIMA1998691}
K.~Tsushima, K.~Saito, J.~Haidenbauer, and A.~Thomas, ``The quark-meson
  coupling model for $\lambda$, $\sigma$ and $\xi$ hypernuclei,'' {\em Nuclear
  Physics A}, vol.~630, no.~3, pp.~691--718, 1998.

\bibitem{Motta:2019tjc}
T.~F. Motta, A.~M. Kalaitzis, S.~Anti\'c, P.~A.~M. Guichon, J.~R. Stone, and
  A.~W. Thomas, ``{Isovector Effects in Neutron Stars, Radii and the GW170817
  Constraint},'' {\em Astrophys. J.}, vol.~878, no.~2, p.~159, 2019.

\bibitem{Motta:2020xsg}
T.~F. Motta, P.~A.~M. Guichon, and A.~W. Thomas, ``{On the sound speed in
  hyperonic stars},'' {\em Nucl. Phys. A}, vol.~1009, p.~122157, 2021.

\bibitem{LIGOScientific:2017vwq}
B.~P. Abbott {\em et~al.}, ``{GW170817: Observation of Gravitational Waves from
  a Binary Neutron Star Inspiral},'' {\em Phys. Rev. Lett.}, vol.~119, no.~16,
  p.~161101, 2017.

\bibitem{LIGOScientific:2018cki}
B.~P. Abbott {\em et~al.}, ``{GW170817: Measurements of neutron star radii and
  equation of state},'' {\em Phys. Rev. Lett.}, vol.~121, no.~16, p.~161101,
  2018.

\bibitem{Riley:2021pd}
T.~E. Riley {\em et~al.}, ``{A NICER View of the Massive Pulsar PSR J0740+6620
  Informed by Radio Timing and XMM-Newton Spectroscopy},'' {\em Astrophys. J.
  Lett.}, vol.~918, no.~2, p.~L27, 2021.

\bibitem{CIARCELLUTI201119}
P.~Ciarcelluti and F.~Sandin, ``Have neutron stars a dark matter core?,'' {\em
  Physics Letters B}, vol.~695, no.~1, pp.~19--21, 2011.

\bibitem{Sandin_2009}
F.~Sandin and P.~Ciarcelluti, ``Effects of mirror dark matter on neutron
  stars,'' {\em Astroparticle Physics}, vol.~32, p.~278–284, Dec 2009.

\bibitem{PhysRevD.77.043515}
G.~Bertone and M.~Fairbairn, ``Compact stars as dark matter probes,'' {\em
  Phys. Rev. D}, vol.~77, p.~043515, Feb 2008.

\bibitem{Mukhopadhyay_2017}
S.~Mukhopadhyay, D.~Atta, K.~Imam, D.~N. Basu, and C.~Samanta, ``Compact
  bifluid hybrid stars: hadronic matter mixed with self-interacting fermionic
  asymmetric dark matter,'' {\em The European Physical Journal C}, vol.~77, Jul
  2017.

\bibitem{Leung:2011zz}
S.~C. Leung, M.~C. Chu, and L.~M. Lin, ``{Dark-matter admixed neutron stars},''
  {\em Phys. Rev. D}, vol.~84, p.~107301, 2011.

\bibitem{Ellis_2018}
J.~Ellis, G.~Hütsi, K.~Kannike, L.~Marzola, M.~Raidal, and V.~Vaskonen, ``Dark
  matter effects on neutron star properties,'' {\em Physical Review D},
  vol.~97, Jun 2018.

\bibitem{Kouvaris_2012}
C.~Kouvaris, ``Limits on self-interacting dark matter from neutron stars,''
  {\em Physical Review Letters}, vol.~108, May 2012.

\bibitem{2021_w}
W.~Husain and A.~W. Thomas, ``Possible nature of dark matter,'' {\em Journal of
  Cosmology and Astroparticle Physics}, vol.~2021, p.~086, Oct 2021.

\bibitem{2000NuPhB.564..185M}
E.~W. {Mielke} and F.~E. {Schunck}, ``{Boson stars: alternatives to primordial
  black holes?},'' {\em Nuclear Physics B}, vol.~1, pp.~185--203, Jan. 2000.

\bibitem{berryman2022neutron}
J.~M. Berryman, S.~Gardner, and M.~Zakeri, ``Neutron stars with baryon number
  violation, probing dark sectors,'' 2022.

\bibitem{2019_reddy}
C.~Horowitz and S.~Reddy, ``Gravitational waves from compact dark objects in
  neutron stars,'' {\em Physical Review Letters}, vol.~122, Feb 2019.

\bibitem{2013_red}
B.~Bertoni, A.~E. Nelson, and S.~Reddy, ``Dark matter thermalization in neutron
  stars,'' {\em Physical Review D}, vol.~88, Dec 2013.

\bibitem{deLavallaz:2010wp}
A.~de~Lavallaz and M.~Fairbairn, ``{Neutron Stars as Dark Matter Probes},''
  {\em Phys. Rev. D}, vol.~81, p.~123521, 2010.

\bibitem{Sen:2021wev}
D.~Sen and A.~Guha, ``{Implications of feebly interacting dark sector on
  neutron star properties and constraints from GW170817},'' {\em Mon. Not. Roy.
  Astron. Soc.}, vol.~504, no.~3, p.~3, 2021.

\bibitem{PhysRevD.105.023001}
D.~Rafiei~Karkevandi, S.~Shakeri, V.~Sagun, and O.~Ivanytskyi, ``Bosonic dark
  matter in neutron stars and its effect on gravitational wave signal,'' {\em
  Phys. Rev. D}, vol.~105, p.~023001, Jan 2022.

\bibitem{Guichon:2008zz}
P.~A.~M. Guichon, A.~W. Thomas, and K.~Tsushima, ``{Binding of hypernuclei in
  the latest quark-meson coupling model},'' {\em Nucl. Phys. A}, vol.~814,
  pp.~66--73, 2008.

\bibitem{Husain:2022bxl}
W.~Husain, T.~F. Motta, and A.~W. Thomas, ``{Consequences of neutron decay
  inside neutron stars},'' 3 2022.

\bibitem{Guichon:1987jp}
P.~A.~M. Guichon, ``{A Possible Quark Mechanism for the Saturation of Nuclear
  Matter},'' {\em Phys. Lett. B}, vol.~200, pp.~235--240, 1988.

\bibitem{PhysRev.55.374}
J.~R. Oppenheimer and G.~M. Volkoff, ``On massive neutron cores,'' {\em Phys.
  Rev.}, vol.~55, pp.~374--381, Feb 1939.

\bibitem{Tolman169}
R.~C. Tolman, ``Effect of inhomogeneity on cosmological models,'' {\em
  Proceedings of the National Academy of Sciences}, vol.~20, no.~3,
  pp.~169--176, 1934.

\bibitem{Hinderer_2008}
T.~Hinderer, ``Tidal love numbers of neutron stars,'' {\em The Astrophysical
  Journal}, vol.~677, p.~1216–1220, Apr 2008.

\bibitem{Hinderer_2010}
T.~Hinderer, B.~D. Lackey, R.~N. Lang, and J.~S. Read, ``Tidal deformability of
  neutron stars with realistic equations of state and their gravitational wave
  signatures in binary inspiral,'' {\em Physical Review D}, vol.~81, Jun 2010.

\bibitem{Abbott_2017}
B.~Abbott, R.~Abbott, T.~Abbott, F.~Acernese, K.~Ackley, C.~Adams, T.~Adams,
  P.~Addesso, R.~Adhikari, V.~Adya, and et~al., ``Gw170817: Observation of
  gravitational waves from a binary neutron star inspiral,'' {\em Physical
  Review Letters}, vol.~119, Oct 2017.

\bibitem{Abbott_2019}
B.~Abbott, R.~Abbott, T.~Abbott, S.~Abraham, F.~Acernese, K.~Ackley, C.~Adams,
  R.~Adhikari, V.~Adya, C.~Affeldt, and et~al., ``Gwtc-1: A gravitational-wave
  transient catalog of compact binary mergers observed by ligo and virgo during
  the first and second observing runs,'' {\em Physical Review X}, vol.~9, Sep
  2019.

\bibitem{Bramante_2018_31}
J.~Bramante, T.~Linden, and Y.-D. Tsai, ``Searching for dark matter with
  neutron star mergers and quiet kilonovae,'' {\em Phys. Rev. D}, vol.~97,
  p.~055016, Mar 2018.

\end{thebibliography}
\bibliographystyle{ieeetr}
\end{document}